\documentclass[pra,nopacs,superscriptaddress,twocolumn,twoside]{revtex4}

\usepackage{amsfonts}
\usepackage{amsmath}
\usepackage{hyperref}
\usepackage{scalefnt}
\usepackage{latexsym}
\usepackage{revsymb}
\usepackage{graphicx}

\newtheorem{theorem}{Theorem}

\newtheorem{proposition}[theorem]{Proposition}
\def\qed{{\ \rule{0.5em}{0.5em}}}
\newenvironment{proof}[1][Proof]{\noindent\textbf{#1.} }{\ \rule{0.5em}{0.5em}}

\renewcommand{\>}{\rangle}

\input{tcilatex}

\newcommand{\qchrom}{{\chi_q}}
\newcommand{\qqchrom}[1]{{\chi_q^{(#1)}}}

\def\squareforqed{\hbox{\rlap{$\sqcap$}$\sqcup$}}
\def\qed{\ifmmode\squareforqed\else{\unskip\nobreak\hfil
\penalty50\hskip1em\null\nobreak\hfil\squareforqed
\parfillskip=0pt\finalhyphendemerits=0\endgraf}\fi}
\def\endenv{\ifmmode\;\else{\unskip\nobreak\hfil
\penalty50\hskip1em\null\nobreak\hfil\;
\parfillskip=0pt\finalhyphendemerits=0\endgraf}\fi}

\mathchardef\ordinarycolon\mathcode`\:
\mathcode`\:=\string"8000
\def\vcentcolon{\mathrel{\mathop\ordinarycolon}}
\begingroup \catcode`\:=\active
  \lowercase{\endgroup
  \let :\vcentcolon
  }

\newcommand{\nc}{\newcommand}
\nc{\rnc}{\renewcommand}
\nc{\beq}{\begin{equation}}
\nc{\eeq}{{\end{equation}}}
\nc{\beqa}{\begin{eqnarray}}
\nc{\eeqa}{\end{eqnarray}}
\nc{\lbar}[1]{\overline{#1}}
\nc{\bra}[1]{\langle#1|}
\nc{\ket}[1]{|#1\rangle}
\nc{\ketbra}[2]{|#1\rangle\!\langle#2|}
\nc{\braket}[2]{\langle#1|#2\rangle}
\nc{\proj}[1]{| #1\rangle\!\langle #1 |}
\nc{\avg}[1]{\langle#1\rangle}
\rnc{\max}{\operatorname{max}}
\rnc{\min}{\operatorname{min}}
\rnc{\inf}{\operatorname{inf}}
\nc{\Rank}{\operatorname{Rank}}
\nc{\smfrac}[2]{\mbox{$\frac{#1}{#2}$}}
\nc{\tr}{\operatorname{Tr}}
\nc{\ox}{\otimes}
\nc{\dg}{\dagger}
\nc{\dn}{\downarrow}
\nc{\cA}{{\cal A}}
\nc{\cB}{{\cal B}}
\nc{\cC}{{\cal C}}
\nc{\cD}{{\cal D}}
\nc{\cE}{{\cal E}}
\nc{\cF}{{\cal F}}
\nc{\cG}{{\cal G}}
\nc{\cH}{{\cal H}}
\nc{\cI}{{\cal I}}
\nc{\cJ}{{\cal J}}
\nc{\cK}{{\cal K}}
\nc{\cL}{{\cal L}}
\nc{\cM}{{\cal M}}
\nc{\cN}{{\cal N}}
\nc{\cO}{{\cal O}}
\nc{\cP}{{\cal P}}
\nc{\cR}{{\cal R}}
\nc{\cS}{{\cal S}}
\nc{\cT}{{\cal T}}
\nc{\cX}{{\cal X}}
\nc{\cZ}{{\cal Z}}
\nc{\csupp}{{\operatorname{csupp}}}
\nc{\qsupp}{{\operatorname{qsupp}}}
\nc{\var}{\operatorname{var}}
\nc{\rar}{\rightarrow}
\nc{\lrar}{\longrightarrow}
\nc{\polylog}{\operatorname{polylog}}
\nc{\1}{{\openone}}

\nc{\RR}{{{\mathbb R}}}
\nc{\CC}{{{\mathbb C}}}
\nc{\FF}{{{\mathbb F}}}
\nc{\NN}{{{\mathbb N}}}
\nc{\ZZ}{{{\mathbb Z}}}
\nc{\PP}{{{\mathbb P}}}
\nc{\QQ}{{{\mathbb Q}}}
\nc{\UU}{{{\mathbb U}}}
\nc{\EE}{{{\mathbb E}}}
\nc{\id}{{\operatorname{id}}}

\nc{\be}{\begin{equation}}
\nc{\ee}{{\end{equation}}}
\nc{\bea}{\begin{eqnarray}}
\nc{\eea}{\end{eqnarray}}
\nc{\Hom}[2]{\mbox{Hom}(\CC^{#1},\CC^{#2})}
\nc{\rU}{\mbox{U}}

\nc{\ob}[1]{#1}

\begin{document}

\title{On the quantum chromatic number of a graph}

\author{Peter J. Cameron}

\author{Michael W. Newman}
\email{m.newman@qmul.ac.uk}
\affiliation{School of Mathematical Sciences, Queen Mary, University of London, London E1 4NS, U.K.}

\author{Ashley Montanaro}
\affiliation{Department of Computer Science, University of Bristol, Bristol BS8 1UB, U.K.}

\author{Simone Severini}
\email{ss54@york.ac.uk}
\affiliation{Department of Mathematics, University of York, York YO10 5DD, U.K.}

\author{Andreas Winter}
\affiliation{Department of Mathematics, University of Bristol, Bristol BS8 1TW, U.K.}


\begin{abstract}
  We investigate the notion of quantum chromatic number of a graph,
  which is the minimal number of colours necessary in a protocol in which
  two separated provers can convince an interrogator with certainty
  that they have a colouring of the graph.

  After discussing this notion from first principles, we go on to establish
  relations with the clique number and orthogonal representations of the
  graph. We also prove several general facts about this graph parameter
  and find large separations between the clique number and the quantum
  chromatic number by looking at random graphs.
  Finally, we show that there can be no separation between classical and
  quantum chromatic number if the latter is $2$, nor if it is $3$ in a
  restricted quantum model; on the other hand, we exhibit a graph on
  $18$ vertices and $44$ edges with chromatic number $5$ and quantum
  chromatic number $4$.
\end{abstract}

\date{13 September 2006}

\maketitle

\section{Introduction}
\label{sec:setting}
Alice and Bob want to convince a referee
with probability $1$ that they have a $c$-colouring of a graph $G=(V,E)$ in
the interrogation model: they each get asked a vertex $v$, $w$ of the graph,
respectively and have to report back a colour $\alpha $, $\beta $ (resp.) to
the referee (\emph{i.e.}, a number in $[c]=\{0,1,\ldots,c-1\}$; if $v=w$,
then to pass they have to reply the same: $\alpha =\beta $; if $vw\in E$,
then to pass they have to reply differently: $\alpha \neq \beta $.
In this paper, a graph will always be undirected and without loops,
i.e.~$E \subseteq {V \choose 2}$.

If they are not allowed to talk to each other during the interrogation
but may agree on a strategy beforehand, then it is straightforward to
see that they can pass with probability $1$ if and only if $c \geq \chi(G)$,
the chromatic number of $G$ -- that is, in a classical world where Alice
and Bob may share randomness and an otherwise deterministic strategy.
However, if Alice and Bob share an entangled state (possibly depending
on the graph), there are graphs for which Alice and Bob can win this game
with probability $1$ for $c < \chi(G)$. Based 
on a suggestion of one of the authors (also, independently of
Patrick Hayden, see~\cite{av}) we call the smallest
$c$ such that Alice and Bob can win the graph colouring game
the \emph{quantum chromatic number}.

Such a problem was first considered in~\cite{BCW,cost-of-e}, and generalised
in~\cite{deWolf:thesis}, Theorems 8.5.1-3, and~\cite{BCWW}, for Hadamard graphs:
the vertices are $n$-bit strings, and two of them are joined by an edge if and only
if their Hamming distance is $n/2$. In these references it is shown that
the game can be won with $c=n$ colours.
This line of investigation was carried further under the heading
``pseudotelepathy'' in~\cite{pseudotele-3,pseudotele-4,pseudotele,av}.
Earlier work of Frankl and R\"odl~\cite{FR} in extremal combinatorics
established that the chromatic number of the Hadamard graphs grows
exponentially in $n$.  In~\cite{mwn} it is shown that the chromatic
number is equal to $n$ if and only if $n \in \{1,2,4,8\}$.

The rest of the paper is structured as follows: in section~\ref{sec:models}
we present the model, or actually an infinite hierarchy of models
for the quantum chromatic number. Then we go on to general properties
of the quantum chromatic number in section~\ref{sec:general},
bounds via orthogonal representations (section~\ref{sec:orth}),
small number of colours (section~\ref{sec:few}), and finally
random graphs (section~\ref{sec:random}), after which we conclude
with a number of open questions and conjectures.

\section{Model(s)}
\label{sec:models}
The most general strategy for Alice and Bob to win the graph colouring
game with probability $1$ with $c$ colours
for a graph $G=(V,E)$ consists of an entangled
state $\ket{\psi}_{AB} \in \CC^{d\times d}$ 
shared between them, and two families of
POVMs $(E_{v\alpha})_{\alpha=0,\ldots,c-1}$ and
$(F_{v\beta})_{\beta=0,\ldots,c-1}$, indexed
by the vertices $v\in V$ of the graph.
The fact that they win with probability $1$ is expressed by the
consistency condition
\begin{equation}\begin{split}
  \label{eq:consistency}
  \forall v\in V\  \forall\alpha\neq\beta &\quad
         \bra{\psi} E_{v\alpha} \ox F_{v\beta} \ket{\psi} = 0, \\
  \forall vw\in E\ \forall\alpha          &\quad
         \bra{\psi} E_{v\alpha} \ox F_{w\alpha} \ket{\psi} = 0.
\end{split}\end{equation}
Note that the dimension $d$ bears no relationship to $c$, that the
entangled state $\psi$ can be anything (it may even be mixed but
it is immediate that w.l.o.g.~we may assume it to be pure),
and the POVMs may have operators of arbitrary rank.

The smallest possible $c$ for which Alice and Bob can convince the referee,
i.e.~such that eq.~(\ref{eq:consistency}) holds,
is called the \emph{quantum chromatic number} of $G$ and it will be denoted
by $\chi_{q}(G)$.

\begin{proposition}
  \label{prop:normalform}
  To win the graph colouring game in the above setting, w.l.o.g.~the state
  is maximally entangled, and the POVM elements are all projectors,
  all w.l.o.g.~of the same rank.
\end{proposition}
\begin{proof}
  Without loss of generality we can assume that $\ket{\psi}$ has full
  Schmidt rank $d$ since otherwise we restrict all POVMs to 
  the supports of the respective reduced states.
  From eq.~(\ref{eq:consistency}) we get, for any $v\in V$, any $\alpha$
  and $\beta\neq\alpha$,
  that $E_{v\alpha} \perp \tr_B \bigl( (\1\ox F_{v\beta}) \proj{\psi} \bigr)$, hence
  \[\begin{split}
    E_{v\alpha} &\perp \sum_{\beta\neq\alpha}
                         \tr_B \bigl( (\1\ox F_{v\beta}) \proj{\psi} \bigr)    \\
                &\phantom{==}
                 =       \tr_B \bigl( (\1\ox\1-\1\ox F_{v\alpha}) \proj{\psi} \bigr).
  \end{split}\]
  From this, and because Alice needs to get outcome $\alpha$ with certainty
  if Bob gets $\alpha$, we must have
  \[
    E_{v\alpha} = {\rm supp}\, \tr_B \bigl( (\1\ox F_{v\alpha}) \proj{\psi} \bigr).
  \]
  By the same argument all $F_{v\beta}$ are projectors.

  Now we argue that the consistency requirement for the state $\psi$ implies
  that it is also true when we substitute the maximally entangled
  state $\Phi_d$: in its Schmidt basis,
  $\ket{\psi} = \sum_i \sqrt{\lambda_i}\ket{i}\ket{i}$, and denoting
  $\rho = \tr_B \proj{\psi} = \sum_i \lambda_i \proj{i} = \tr_A \proj{\psi}$,
  the finding of the previous paragraph can be cast as
  \begin{align*}
    E_{v\alpha} &= {\rm supp}\, \sqrt{\rho} \bigl(\overline{F_{v\alpha}}\bigr) \sqrt{\rho}, \\
    F_{w\beta}  &= {\rm supp}\, \sqrt{\rho} \bigl(\overline{E_{w\beta}}\bigr) \sqrt{\rho}.
  \end{align*}  
  This implies however
  \[
    E_{v\alpha}\rho E_{v\beta} = 0
  \]
  for all $v$ and $\alpha\neq\beta$ (where we cancelled $\sqrt{\rho}$'s
  left and right), and likewise for $F_{w\alpha},\, F_{w\beta}$.
  But with the fact that each $E_{v\alpha}$
  is a projector and that summed over $\alpha$ they yield the identity,
  this gives (for arbitrary $v$)
  \[
    \rho = \sum_{\alpha,\beta} E_{v\alpha}\rho E_{v\beta}
         = \sum_\alpha E_{v\alpha}\rho E_{v\alpha},
  \]
  from which it follows that $\rho$ commutes with all the (Kraus) operators
  $E_{v\alpha}$, and likewise $F_{w\beta}$~\cite{lindblad}. Hence we find
  \[
    E_{v\alpha} = \overline{F_{v\alpha}},\quad 
    F_{w\beta}  = \overline{E_{w\beta}},
  \]
  and that is the claim we set out to prove: we may as well
  assume that $\ket{\psi}$ is maximally entangled.

  Finally, how to make the operators all the same rank: let
  $\ket{\psi'} = \ket{\psi}\ox\ket{\Phi_c}$, and
  \begin{align*}
    E_{v\alpha}' &:= \sum_{i=0}^{c-1} E_{v,\alpha+i}\ox\proj{i}, \\
    F_{w\beta}'  &:= \sum_{i=0}^{c-1} F_{w,\beta+i} \ox\proj{i},
  \end{align*}
  where the colours are w.l.o.g.~$\{0,\ldots,c-1\}$ and the additions
  above are modulo $c$.
  These states and operators evidently still make for a valid
  quantum colouring, and also clearly all operators
  have now the same rank.
\end{proof}

\medskip
This proposition motivates us to introduce rank-$r$ versions of the
quantum chromatic number:
$\qqchrom{r}(G)$ is the minimum $c$ such that Alice and Bob can win the
graph colouring game for $G$ with a maximally entangled state of rank $rc$,
and POVMs with operators of rank $r$ (exactly).
Then it is clear that $\qqchrom{r}(G) \leq \qqchrom{s}(G)$
whenever $r \geq s$, and that $\qchrom(G) = \inf_r\{\qqchrom{r}(G)\}$.

The special case of rank-$1$ model is the following: 
Alice and Bob share a $c$-dimensional maximally entangled state
\begin{equation*}
  \ket{\Phi_{c}} = \frac{1}{\sqrt{c}}\sum_{i=0}^{c-1} \ket{i}_A \ket{i}_B.
\end{equation*}
To make their choices, they both use rank-$1$ von Neumann measurements,
which are ordered bases $(\ket{e_{v\alpha}})_{\alpha}$ and
$(\ket{f_{v\beta}})_{\beta}$ for all vertices $v$, for Alice
and Bob, respectively.

\medskip\noindent
\textbf{Observation 1.}
Bob's bases are tied to Alice's by the
demand of consistency: we need, for all $v$ and $\alpha $, 
\begin{equation*}
  \bra{e_{v\alpha}}\bra{f_{v\alpha}} \Phi _{c}\rangle = 1/c,
\end{equation*}
which enforces 
\begin{equation*}
  \ket{f_{v\alpha}} = \overline{\ket{e_{v\alpha}}}.
\end{equation*}

\noindent
\textbf{Observation 2.}
This means that we can translate the
colouring condition into something that only concerns Alice's bases: we
need, for all $vw\in E$ and all $\alpha$, 
\begin{equation*}
  \bra{e_{v\alpha}}\bra{f_{w\alpha}} \Phi_{c} \rangle = 0.
\end{equation*}
Because of 
\begin{equation*}
  \bra{f_{w\alpha}} \Phi _{c} \rangle = \frac{1}{\sqrt{c}}\overline{\ket{f_{w\alpha}}}
\end{equation*}
and Observation 1 this can be rewritten as 
\begin{equation}
  \forall vw\in E\text{ and }\forall \alpha \quad
    \bra{e_{v\alpha}} e_{w\alpha}\rangle = 0.
  \label{eq:condition}
\end{equation}

\noindent
\textbf{Observation 3.}  It is convenient to introduce unitary
matrices $U_{v}$ for each vertex $v$, whose columns are just the
vectors $|e_{v\alpha }\rangle $, $\alpha = 0,\ldots,c-1$. Then we can
reformulate Alice's strategy as follows: on receiving the request for
vertex $v$, she performs the unitary $U_{v}^{\dagger}$ on her quantum
system and measures in the standard basis to get a number $\alpha \in
\lbrack c]$. By Observation 1 above, Bob, for vertex $w$, performs the
unitary $\overline{U_{w}}^{\dagger }=U_{w}^{\top }$ and measures in
the standard basis to obtain $\beta \in \lbrack c]$. In the light of
Observation 2, we can rewrite the colouring condition expressed in
eq.~(\ref{eq:condition}) as:
\begin{equation}
  \forall vw\in E\quad U_{v}^{\dagger} U_{w}\text{ has only zeroes on the diagonal.}
  \label{eq:unitaries}
\end{equation}

By a similar chain of arguments we can show, for the POVM constructed
in the proof of proposition~\ref{prop:normalform}, that
$F_{v\alpha} = \overline{E_{v\alpha}}$ for all vertices $v$ and all
colours $\alpha$, and that hence the colouring condition can
be phrased entirely in terms of Alice's operators:
\begin{equation}
  \label{eq:op-consistency}
  \forall vw\in E\text{ and }\forall \alpha \quad E_{v\alpha}E_{w\alpha} = 0,
\end{equation}
i.e.~$E_{v\alpha}$ and $E_{w\alpha}$ are orthogonal.

\section{General Properties}
\label{sec:general}

We look at some basic properties of the quantum chromatic number as a
graph parameter.  None of these are particularly surprising; indeed
the point of this section is to show that the quantum chromatic number
``does the right thing'', and merits being considered as a
generalization of the (ordinary) chromatic number.

A \emph{homomorphism} is a mapping from one graph to another that
preserves edges.  That is, a homomorphism $\phi$ from $G$ to $H$ maps
vertices of $G$ to vertices of $H$ such that if $x$ and $y$ are
adjacent in $G$ then $\phi(x)$ and $\phi(y)$ are adjacent in $H$.  We
write $G \to H$ to indicate that there exists a homomorphism from $G$
to $H$.

The following easy observation is a useful tool.
\begin{proposition}\label{prop:hom}
  If $G \to H$, then $\qqchrom{r}(G) \leq \qqchrom{r}(H)$ for all $r$
  and hence $\qchrom(G) \leq \qchrom(H)$.
\end{proposition}
\begin{proof}
  Let $\phi$ be a homomorphism from $G$ to $H$.  Then any quantum
  colouring of $H$ gives a quantum colouring of $G$ by colouring the
  vertex $x$ of $G$ with the colour assigned to $\phi(x)$ in $H$.
\end{proof}

\medskip
It is trivial to see that if (and only if) $G$ has no edges then
$\qqchrom{r}(G)=\qchrom(G)=1$.  With a little more effort, one sees
that if $G=K_n$ then $\qqchrom{r}(G)=\qchrom(G)=n$.  For, using
proposition~\ref{prop:normalform} and eq.~(\ref{eq:op-consistency}),
we have a set of $n$ rank-$r$ pairwise orthogonal operators in a space
of dimension $cr$.  We can say a little more.
\begin{proposition}\label{prop:bip}
  $\qchrom(G) = 2$ if and only if $\chi(G)=2$.
\end{proposition}
\begin{proof}
  If $\chi(G)=2$, then $G \to K_2$ and $K_2 \to G$, and so by
  proposition~\ref{prop:hom} $\qchrom(G)$ is at most and at least $2$.
\end{proof}

\medskip
The \emph{clique number of $G$}, denoted by $\omega(G)$ is the size of
the largest complete subgraph of $G$.
\begin{proposition}\label{prop:qineqs}
  $\omega(G) \leq \qchrom(G) \leq \chi(G)$ \qed
\end{proposition}
\begin{proof}
  Any graph $G$ contains $K_{\omega(G)}$ as a subgraph, hence
  $K_{\omega(G)} \to G$.  Also $G \to K_{\chi(G)}$, by mapping each
  vertex to the vertex of $K_{\chi(G)}$ corresponding to its colour.
  The result follows by proposition~\ref{prop:hom}.
\end{proof}

\medskip
Of course, propositions~\ref{prop:bip}~and~\ref{prop:qineqs} remain
valid if we replace $\qchrom$ with $\qqchrom{r}$ for any $r$.

\medskip
Let $G$ and $H$ be two graphs on the same vertex set.  We define the
graph $G \cup H$ to be the graph whose edge set is the union of the
edge sets of $G$ and $H$.  It is a well-known result in graph
theory~\cite[Chap~14.1]{ore} that $\chi(G \cup H) \leq \chi(G)
\chi(H)$: colour each vertex in $G \cup H$ with the ordered pair of
colours it received in colourings of $G$ and $H$, respectively.  This
idea can be extended to quantum colourings:
\begin{proposition}\label{prop:chiprod}
  For any $r,s$, we have $\qqchrom{rs}(G \cup H) \leq \qqchrom{r}(G)\qqchrom{s}(H)$.
\end{proposition}
\begin{proof}
  Given rank-$r$ and rank-$s$ quantum colourings for $G$ and $H$
  respectively, we obtain a rank-$rs$ quantum colouring of $G \cup H$
  by taking the tensor products of the individual POVM operators
  associated to the vertices.
\end{proof}

\medskip
As a corollary, we obtain the following, showing that a graph and its
complement cannot both have small quantum chromatic number.
\begin{proposition}
  $\qchrom(G)\qchrom(\overline{G}) \geq n$.
\end{proposition}
\begin{proof}
  Apply proposition~\ref{prop:chiprod} with $H=\overline{G}$, the
  complement of $G$.
\end{proof}

\section{Orthogonal Representations}
\label{sec:orth}

The origin of the quantum chromatic number is in Hadamard
graphs~\cite{BCW,cost-of-e}, which are a special case of orthogonality
graphs, so it is natural to consider the larger family.

An \emph{orthogonal representation} of a graph $G$ is a mapping
$\phi$ from the vertices of $G$ to the non-zero vectors of some
vector space, such that if two vertices $x$ and $y$ are adjacent, then
$\phi(x)$ and $\phi(y)$ are orthogonal.

Given a set of vectors, we define their \emph{orthogonality graph} to
be the graph having the vectors as vertices, with two vectors adjacent
if and only if they are orthogonal.

Let $\xi(G)$ to be the smallest integer $c$ such that $G$ has an
orthogonal representation in the vector space $\CC^c$.
Furthermore, let $\xi'(G)$ to be the smallest integer $c$ such that
$G$ has an orthogonal representation in the vector space $\CC^c$
with the added restriction that the entries of each vector must have
modulus one.  (Note that we really only need the entries in any
particular vector to have constant modulus.)

\begin{proposition}\label{prop:xineqs}
  $\omega(G) \leq \xi(G) \leq \qqchrom{1}(G) \leq \xi'(G) \leq \chi(G)$
\end{proposition}
\begin{proof}
  For each integer $c$, let $F_c$ be the discrete Fourier transform of
  order $c$, i.e., $[F_{c}]_{j,k}=\frac{1}{\sqrt{c}}e^{2\pi ijk/c}$.

  Three of these inequalities are straightforward.

  Given a graph with $\chi(G)=c$, colour the vertices with the rows of
  $F$. Adjacent vertices have distinct colours and hence orthogonal
  vectors, and thus $\xi'(G) \leq \chi(G)$.
  Given a graph with $\qqchrom{1}(G)=c$, map each vertex to the first
  column of its corresponding unitary matrix.
  By~eq.~(\ref{eq:condition}) adjacent vertices will get mapped to
  orthogonal vectors, and thus $\xi(G) \leq \qqchrom{1}(G)$.
  Given a graph with $\omega(G)=c$, any orthogonal representation of
  it must contain $c$ pairwise orthogonal vectors and thus $\omega(G)
  \leq \xi(G)$.

  Finally, given a graph with $\xi'(G)=c$, map each vertex $x$ to $\Delta_xF_c$,
  where $\Delta_x$ is the diagonal (unitary) matrix whose diagonal
  entries are the entries of $x$.  Then $\langle x | y \rangle =0 $
  implies that $(\Delta_{v}F_c)^{\dagger }(\Delta _{w}F_c)$ has only
  zeroes on the diagonal.  Thus $\qqchrom{1} \leq \xi'$.
\end{proof}

\medskip
The proof that $\qqchrom{1}(G)
\leq \xi'(G)$ is in fact a familiar one: it is essentially the
original proof of~\cite{BCW,cost-of-e} using $F_c$ in place of a
Hadamard matrix, or extension of~\cite{av} using more general
vertices.

In fact the only properties of $F_c$ that we need are that its columns
form an orthonormal basis and the entries all have the same modulus.
So the (normalized) character table of \emph{any} Abelian group of
order $c$ will do (as will a generalized Hadamard matrix). Likewise,
the only properties of the vertices that we need are that adjacent
vertices are orthogonal and the entries all have the same modulus, so
we need not restrict ourselves to $\pm1$-vectors.

\medskip
The results of~\cite{BCW,cost-of-e} can be rephrased in our current
language as follows.  Given a graph $G$, what is the smallest integer
$c$ such that $G$ has an orthogonal representation in $\CC^c$ with the
added restriction that all entries are $\pm1$.  This motivates us to
consider the following question: what happens if we replace ``$\pm1$''
by some other subset of roots of unity? 

\begin{proposition}\label{prop:prime}
  Let $p$ be a prime.  Let $G$ be the graph whose vertices are the
  vectors of $\CC^p$ whose entries are all $p$-th roots of unity.
  Then $\chi(G)=p$.
\end{proposition}
\begin{proof}
  We first show that $G$ is a Cayley graph for $\ZZ_{p}^{n}$ (this is
  in fact well known).  To each vertex $a$ associate the mapping
  $\sigma_{a} : x \to a \circ x$, where $a \circ x$ denotes the
  entry-wise product of $a$ and $x$.
  Two vertices $x$ and $y$ are adjacent when $\langle x | y \rangle =
  0$, or equivalently when $y=\sigma_{a}(x)$ for some $a$ whose
  entries sum to zero.  Thus the connection set is the set of such
  $a$.

  The fact that $p$ is prime is relevant for the following reason.
  Vertices $x$ and $y$ are adjacent if and only if the entries of
  $\overline{x} \circ y$ are all distinct: this is because the entries
  are $p$-th roots of unity and there are $p$ of them.

  It is well-known that for vertex transitive graphs $H$ (such as
  Cayley graphs), we have $\alpha(H)\omega(H) \leq v$.  We use an
  extension due to Godsil~\cite{cols}, which in our case we may state
  as follows: if $H$ is a Cayley graph on $v$ vertices for an Abelian
  group then $\alpha(H)\omega(H) = v$ if and only if
  $\chi(H)=\omega(H)$.

  It is easy to see that $\omega(G)=p$: take the rows of the character
  table for $\ZZ_{p}$.  So it is necessary and sufficient to find an
  independent set of size $p^{p-1}$.

  The set of vertices $x$ with $x_1=x_2$ form an independent set of
  size $p^{p-1}$: no two of them are adjacent since for any such $x$
  and $y$, the first two entries of $\overline{x} \circ y$ are equal,
  hence the entries are not all distinct.
\end{proof}

\medskip
Note that in an orthogonality graph, vectors that differ by a scalar
multiple are non-adjacent and have the same neighbours, so we may
restrict ourselves to vectors that have first entry equal to one.  (We
are really dealing with $1$-dimensional subspaces and not vectors.)
For convenience, we use this in the next result.

\begin{proposition}
  \label{prop:4-dim}
  Let $G$ be the orthogonality graph defined by vectors of dimension $4$
  whose entries are taken from the set $\{1,i,-1,-i\}$.
  Then $\chi(G)=4$.
\end{proposition}
\begin{proof}
  We give an explicit 4-colouring of $G$ found by computer. Consider the
  set $S$ of all 4-dimensional vectors whose first component is 1, and
  whose other 3 components are taken from the set $\{1,i,-1,-i\}$. A
  4-colouring of the orthogonality graph of $S$ gives a 4-colouring of
  $G$. Consider each element $s \in S$ as a 3-digit string
  $s' \in \{0,1,2,3\}^3$ giving the power of $i$ in each component of
  $s$, and list the vertices of $S$ in lexicographic order.
  Then, a 4-colouring of the orthogonality graph of $S$ is given
  by the following:
  \[
  \begin{split}
    (& 1,1,1,2,1,1,1,2,2,1,2,2,2,1,2,2,2,1,3,3,4,1,1,3,4, \\
     & 4,1,2,2,4,3,2,3,4,3,3,3,4,3,3,4,4,4,3,4,4,4,3,1,1, \\
     & 4,3,3,1,2,3,3,4,2,2,1,4,4,2).
  \end{split}
  \]
\end{proof}

\medskip
Both the graphs of proposition~\ref{prop:prime} and
proposition~\ref{prop:4-dim} satisfy $\omega=\chi$, and therefore by
proposition~\ref{prop:qineqs} also have $\omega=\qchrom=\qqchrom{r}=\chi$.

It is not hard to see directly that the orthogonality graph on
$\pm1$-vectors of dimension $2$ is $2$-colourable.  Thus for
orthogonality graphs using $\pm1$-vectors, in order to have $\qchrom <
\chi$ we need to go to dimensions larger than $2$.  We now see that
for $d$ a prime or $d=4$, using $d$-th roots of unity vectors forces
us to go to dimensions larger than $d$ to obtain $\qchrom<\chi$.

\medskip
Finally, we derive an upper bound on the orthogonality graph of
$\CC^k$, which gives an upper bound on $\chi(G)$ in terms of $\xi(G)$.
This allows us to bound the largest possible gap between $\chi(G)$ and
$\qqchrom{1}(G)$.

\begin{proposition}
  \label{prop:upperbound}
  For any graph $G$,
  \[
    \chi(G) \leq (1+2\sqrt{2})^{2\,\xi(G)} \leq (1+2\sqrt{2})^{2\qqchrom{1}(G)}.
  \]
\end{proposition}
\begin{proof}
  To show the first inequality, we give a colouring of
  the orthogonality graph on $\CC_k$, where $k=\xi(G)$.
  This can be produced from a set of unit vectors $V=\{|v_i\>\}$ such
  that for all unit vectors $|w\> \in \mathbb{C}^k$, $\||w\>-|v_i\>\|_2<1/\sqrt{2}$
  for some $i$, by assigning colour $i$ to $|w\>$ (if there are two or more
  vectors in $V$ satisfying this inequality, picking one arbitrarily). This
  works because $\braket{w}{v_i}=0 \Rightarrow 2(1-\mathrm{Re}(\braket{w}{v_i})) =
  \||w\>-|v_i\>\|_2^2 = 2$, so no two orthogonal vectors will receive the
  same colour. We use the argument of \cite{hayden} to bound the size of such
  a set (which \cite{hayden} calls a $1/\sqrt{2}$-net). Let $M=\{|v_i\>\}$ be a
  maximal set of unit vectors such that $\||v_i\>-|v_j\>\|_2\ge 1/\sqrt{2}$ for
  all $i$ and $j$. Then $M$ is a $1/\sqrt{2}$-net. Set $m=|M|$ and observe that,
  as subsets of $\mathbb{R}^{2k}$, the open balls of radius $1/(2\sqrt{2})$
  about each $|v_i\>$ are disjoint and contained in the overall ball of radius
  $1+1/(2\sqrt{2})$. Thus $m(1/(2\sqrt{2}))^{2k} \le (1+1/(2\sqrt{2}))^{2k}$.
  The second inequality follows from proposition \ref{prop:xineqs}.
\end{proof}

\medskip\noindent
{\bf Remark.}
The above result shows that the separation between $\chi(G)$
and $\qqchrom{1}(G)$ can be at most exponential; the results
of~\cite{BCWW,deWolf:thesis} on the other hand demonstrate
that exponential gaps can occur, showing that this is indeed the
most extreme case, up to a constant factor in the exponent.

\section{Few colours}
\label{sec:few}
Here we investigate properties of graphs with small quantum chromatic
number or small orthogonal rank. We already saw that for two colours,
classical and quantum chromatics numbers coincide. It turns out that
for three this is also the case, and for numbers up to $8$ the quantum
chromatic number stays close to the orthogonal rank.

\begin{proposition}
  \label{prop:small}
  Given a graph $G$, $\qqchrom{1}(G)=3$ if and only if $\chi(G)=3$.
\end{proposition}
\begin{proof}
  If $\chi(G)=3$, we cannot have $\qchrom(G)=2$ (nor $1$ because the
  graph is not empty) as this would mean $\chi(G)=2$. On the other
  hand, consider a rank-$1$ quantum colouring with $3$ colours. We use
  the analysis in section~\ref{sec:models} and in particular the last
  observation 3: we can view the quantum colouring as a family of
  $3\times 3$-unitaries $U_v$ such that eq.~(\ref{eq:unitaries}).  The
  columns of the unitaries are just the basis vectors $\ket{e_{v0}}$,
  $\ket{e_{v1}}$, $\ket{e_{v2}}$. W.l.o.g.~the graph is connected and
  for one distinguished vertex $v_0$ we may assume $U_{v_0} = \1$.

  The crucial observation is that there are essentially only two unitary(!)
  matrices $U_v^\dagger U_w$ with zeroes on the diagonal~\cite{pattern}:
  they can only be
  \[
    \left(\begin{array}{ccc}
            0 & 0 & * \\
            * & 0 & 0 \\
            0 & * & 0
          \end{array}\right)\ \text{ or }\ 
    \left(\begin{array}{ccc}
            0 & * & 0 \\
            0 & 0 & * \\
            * & 0 & 0
          \end{array}\right),
  \]
  where the starred entries must be roots of unity. Starting from
  $v_0$ we hence find inductively that all $U_v$ are, up to phase
  factors, permutation matrices. Just looking at the first column,
  we now obtain a $3$-colouring of $G$, choosing the colour
  according to the row in which the nonzero entry of the column
  vector is.
\end{proof}

\begin{proposition}
  \label{prop:4:8}
  Let $G$ be a graph with an orthogonal representation in $\mathbb{R}^{c}$. If 
  $c=3,4$ then $\qqchrom{1}(G)\leq 4$; if $4<c\leq 8$ then $\qqchrom{1}(G)\leq 8$.
\end{proposition}
\begin{proof}
If $c=4,8$ then associate every vector $v\in \mathbb{R}^{4}$ and $w\in 
\mathbb{R}^{8}$ to real orthogonal designs $V$ and $W$ of the form
$OD\left(4;1,\ldots,1\right)$ and $OD\left( 8;1,\ldots,1\right) $, respectively. For
example, every vector $v\in \mathbb{R}^{4}$ is associated to a
real-orthogonal matrix 
\begin{equation*}
  V=\left(\begin{array}{cccc}
             v_{1} &  v_{2} &  v_{3} &  v_{4}  \\ 
            -v_{2} &  v_{1} & -v_{4} &  v_{3}  \\ 
            -v_{3} &  v_{4} &  v_{1} & -v_{2}  \\ 
            -v_{4} & -v_{3} &  v_{2} &  v_{1}
          \end{array}\right).
\end{equation*}
If $v\in \mathbb{R}^{c}$ and $c=3$ or $4<c\leq 8$ then concatenate a
zero-vector of length $1$ or $8-c$ to $v$, respectively, and proceed as above.
\end{proof}

\medskip\noindent
\textbf{Remark.} The above construction works based on the
fact that in dimensions $4$ and $8$ there exist division algebras
(Hamilton quaternions and Cayley octonions); namely, the generating
orthogonal units $1,i,j,k,\ldots$ have the property that multiplication
by one of them turns every vector into an orthogonal one.
Unfortunately they exist only in dimensions $1$, $2$, $4$ and $8$,
cf.~\cite{numbers}.

\begin{figure}[htp]
\includegraphics{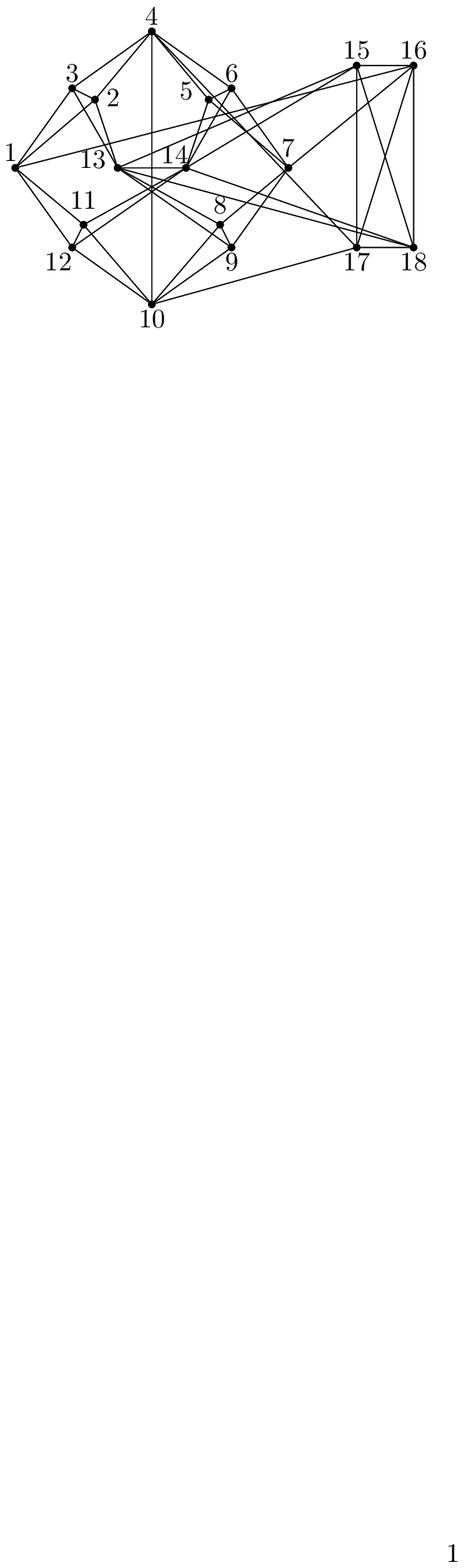}
\caption{A graph $G$ with $\qchrom(G)=\qqchrom{1}(G)=4$, but $\chi(G)=5$.}
\label{fig:smallgraph}
\end{figure}

\medskip\noindent
{\bf Example.}
We now give an example of a fairly small graph $G$ ($18$ vertices and $44$ edges)
which has quantum chromatic number [actually even $\qqchrom{1}(G)$] equal to $4$,
but chromatic number $5$. Label the vertices with integers $1\dots 18$; then
\[
\begin{split}
E=\{ &(1,2),(1,3),(1,11),(1,12),(1,16),(2,3),(2,4),\\
& (2,13),(3,4),(3,13),(4,5),(4,6),(4,10),(4,17),\\
& (5,6),(5,7),(5,14),(6,7),(6,14),(7,8),(7,9),\\
& (7,16),(8,9),(8,10),(8,13),(9,10),(9,13),(10,11),\\
& (10,12),(10,17),(11,12),(11,14),(12,14),(13,14),\\
& (13,15),(13,18),(14,15),(14,18),(15,16),(15,17),\\
& (15,18),(16,17),(16,18),(17,18)\}
\end{split}
\]
The graph may be visualised as consisting of two components connected to each
other by 8 additional edges: a 4-regular graph on vertices $1-14$
[augmented by two edges $(4,10)$ and $(13,14)$], and a $4$-clique on vertices
$15-18$, see Fig.~\ref{fig:smallgraph}.
The following list of vectors gives an orthogonal representation
of $G$ in $\mathbb{R}^{4}$, which by Proposition \ref{prop:4:8} gives a quantum
colouring with 4 colours:
\[
\begin{split}
\{ & (0,0,1,-1),(1,0,0,0),(0,1,1,1),(0,1,0,-1),(0,0,1,0),\\
   & (1,1,0,1),(1,-1,0,0),(0,0,0,1),(1,1,1,0),(1,0,-1,0),\\
   & (0,1,0,0),(1,0,1,1),(0,1,-1,0),(1,0,0,-1),(1,1,1,1),\\
   & (1,1,-1,-1),(1,-1,1,-1),(1,-1,-1,1) \}
\end{split}
\]
Because $G$ contains a $4$-clique, $\qchrom(G)$ cannot, on the other
hand, be smaller than $4$.

It may be verified as follows that $G$ cannot be 4-coloured. Assume
w.l.o.g.\ that vertices 15-18 are coloured $1,2,3,4$ respectively. Then vertices
13 and 14 must divide colours 2 and 3 between them; and for a valid 4-colouring,
none of the triplets $(1,4,13)$, $(1,10,14)$, $(4,7,14)$, $(7,10,13)$ may consist
of 3 distinct colours. Using this, it is straightforward to try all the possible
colourings of vertex 7 and see that each leads to vertices 4 and 10 being assigned
the same colour.

This graph is much smaller and uses fewer colours than the previously
smallest specimen exhibiting a separation between classical and quantum
chromatic numbers: in~\cite{av} a graph on 1609 vertices is described
with $\chi(G)\geq 13$ and $\qchrom(G)=12$.

\section{Random graph properties}
\label{sec:random}
Now we show, in contrast to all previous constructions separating classical
and quantum chromatic number, that the clique number and the quantum
chromatic number (in the rank-$1$ model) are generically exponentially
separated. We use the customary notation ${\cal G}(n,p)$ for the
family of graphs on $n$ vertices with all edges drawn independently
with probability $p$.

\begin{proposition}
  \label{prop:lovasz}
  For a random graph $G\in{\cal G}(n,p)$, and $\epsilon >0$,
  \begin{align*}
    \omega(G)      &\leq (1+\epsilon)\frac{2\log n}{\log 1/p}, \\
    \qqchrom{1}(G) &\geq (1-\epsilon)c(p)\sqrt{n},
  \end{align*}
  almost surely, with some constant $c(p)$ depending on $p$.
\end{proposition}
\begin{proof}
  The statement on the clique number is Bollob\'{a}s' classic
  result~\cite{bollobas} (we actually use a slightly weaker version).

  The statement on $\qqchrom{1}(G)$ follows from that quantity
  being lower bounded by $\xi(G)$, which is lower bounded
  by the Lov\'{a}sz theta function~\cite{lovasz} of the complement graph $\overline{G}$,
  whose random graph behaviour has been worked out by Juhasz~\cite{juhasz}.
\end{proof}

\medskip\noindent
{\bf Remark.} From proposition~\ref{prop:chiprod} we know that for any
graph $G$, $\qchrom(G)\qchrom(\overline{G}) \geq \qchrom(K_n) = n$,
hence at least one of $G$ and $\overline{G}$ has quantum chromatic
number $\geq \sqrt{n}$.
Assume that $G\in{\cal G}(n,1/2)$; then also $\overline{G}\in{\cal G}(n,1/2)$
and both $G$ and $\overline{G}$ are likely to have clique number
only $2\log n+o(\log n)$. That means that we get an abundance of graphs for
which the quantum chromatic number is exponentially larger than the
clique number; asymptotically at least half of all graphs have this
property. It would be interesting to see if the gap between $\omega(G)$
and $\qchrom(G)$ cannot be larger than exponential, in 
extension of proposition~\ref{prop:upperbound}.

\section{Conclusions \&{} Conjectures}
\label{sec:conclusion}
We have studied the quantum chromatic number, the minimal number
of colours required for two independent provers to win the graph
colouring game if they are allowed entanglement, from first principles,
and as a general graph property, beyond the immediate interest of
quantum advantage exhibited in pseudo-telepathy.

We discovered a number of relations of this graph quantity to
other, known, quantities such as chromatic number, clique number,
orthogonal representations and the Lov\'{a}sz theta.
We also found several separations between the quantum chromatic numbers
and these quantities, but had to leave open a number of
important questions.

One of them is the fundamental one: whether the graph colouring game can
always be won with minimal $c$ and a rank-$1$ measurement, in other words,
whether $\qqchrom{1}(G)=\qchrom(G)$ for all graphs $G$.
This has bearing on the decidability of the quantum chromatic number:
the problem if $\qqchrom{r}(G) \leq c$ is decidable because it
boils down to solving the set of quadratic equations~(\ref{eq:consistency})
over the reals in a space of dimension $cr$, for which
there exist exact algorithms based on extensions of
the Gr\"obner basis technique~\cite{real-AG}.
However, $\qchrom(G) = \inf_{r} \qqchrom{r}(G)$
is not decidable in such an easy way. It should be possible to prove at
least an upper bound on $r$ that is sufficient to attain the limit.
In that case, it would make sense to ask about the complexity
of computing $\qchrom(G)$, in particular whether it is NP-hard,
as is computing the chromatic number $\chi(G)$.

Similarly, we found an exponential upper bound on $\chi(G)$ in terms of
$\qqchrom{1}(G)$, but not in terms of $\qchrom(G)$. In particular, it is
still open whether there exists an (infinite) graph $G$ with $\chi(G)=\infty$
and finite $\qchrom(G)$.

Related to the question of whether $\qqchrom{1}(G)=\qchrom(G)$ is the
question of separating $\xi(G)$ and $\xi'(G)$.  If in fact these two
parameters are equal for all graphs, then the rank-$1$ quantum
chromatic number is exactly the minimum dimension for which the graph
has an orthogonal representation.

An interesting question arises in the random graph setting: what is
the likely quantum chromatic number of $G\in{\cal G}(n,p)$?
Conjecture: random graphs have $\chi(G) = \qchrom(G)$ almost surely.
Recalling that $\chi(G) \sim \frac{n \log\frac{1}{1-p}}{2\log n}$
with high probability~\cite{bollobas}, it
may be possible to show by an easier approach that for all $\epsilon >0$,
$\qqchrom{1}(G) \geq (1-\epsilon)\frac{n \log\frac{1}{1-p}}{2\log n}$
almost surely. Namely, one would have to show that the consistency
equations~(\ref{eq:condition}) have, with high probability,
no solution for
$c = \left\lfloor (1-\epsilon)\frac{n \log\frac{1}{1-p}}{2\log n} \right\rfloor$
colours.

Finally, an easier but still fascinating problem would be to find
the smallest graph (and the smallest number of colours) exhibiting
a separation between classical and quantum chromatic number. The graph $G$
shown in Figure \ref{fig:smallgraph} has $\qqchrom{1}(G)=4$ and $\chi(G)=5$.
By proposition~\ref{prop:small}, this is the minimum value of $\qqchrom{1}$
that can achieve such a separation. However, a graph showing a separation with
a smaller number of vertices might exist, as might a graph with $\qchrom(G)=3$,
$\chi(G)>3$.

\acknowledgments
The authors thank Harry Buhrman, Matthias Christandl, Sean Clark,
Patrick Hayden and Troy Lee for discussions on various aspects of this
paper; in particular thanks to Ronald de Wolf for his kind remarks on
an earlier version.

MWN is supported by NSERC Canada.
SS acknowledges support by the U.K.~EPSRC.
AM acknowledges support by the U.K.~EPSRC's ``QIP IRC''.
AW was supported by the U.K.~EPSRC's ``QIP IRC'' and the EC project
QAP (contract no.~IST-2005-15848), as well as by a University of Bristol
Research Fellowship.

\end{document}